\documentclass[prl,amsmath,amssymb,aps,floatfix,nofootinbib,twocolumn]{revtex4}
\usepackage{dcolumn}
\usepackage{bm}
\usepackage{color}
\usepackage[dvips]{epsfig}
\usepackage{graphicx} \usepackage{amsmath} \usepackage{amssymb}

\newcommand{\comment}[1]{}

\newcommand{\BEA}{\begin{eqnarray}}
\newcommand{\EEA}{\end{eqnarray}}

\newcommand{\bq}{\begin{equation}}
\newcommand{\eq}{\end{equation}}
\newcommand{\be}{\begin{eqnarray}}
\newcommand{\ee}{\end{eqnarray}}
\newcommand{\ba}{\begin{align}}
\newcommand{\ea}{\end{align}}

\renewcommand{\d}{{\rm d}}

\newcommand{\uno}{{\cal I}}

\newcommand{\eps}{\epsilon}

\newcommand{\la}{\lambda }
\newcommand{\J}{{\cal J} }

\renewcommand{\i}{I}
\renewcommand{\j}{J}

\begin{document}

\title{Classical dissipative search of unstructured database }

\author{A.E. Allahverdyan and Y. Bisharyan}
\affiliation{Alikhanyan National Science Laboratory (Yerevan Physics Institute),\\ 
2 Alikhanyan Brothers street, Yerevan 0036, Armenia}

\begin{abstract} We propose a physical realization of the unstructured
database search that works via classical, dissipative model of spherical
spins. The database is implemented via spin-spin couplings, where the
selected coupling refers to a larger ferromagnetic interaction between
two selected spins. The low-temperature equilibrium of this model leads
to magnetization strongly concentrated on the selected spins, which
means that the search is complete. The search time refers to the
relaxation time to equilibrium from a homogeneous initial state, and is
described via Langevin equations. This time scales as ${\cal
O}(M^a)$ with $a<1/2$, where $M$ is the database volume. This is faster
than Grover's search, showing how a dissipative, classical analog
computer can overcome the quantum unitary computer.

\end{abstract}

\maketitle

\comment{
References to be included:

- Marti computation via quantum thermodynamics
- Thermodynamics of computation https://arxiv.org/pdf/1911.01968

Check again Eqs.(28--31).

Abstract: mentioning the context of the trade-off is not very clear and convenient.
}

Analog computers operate via continuous variables (e.g., magnetization
or optical intensity) and encode the information directly, whereas usual
digital computers work with discrete representations \cite{0,1}.  Analog
computers closely relate to the structure of the system they intend to
simulate: they connect the computational model and the computing
substrate. For centuries, analog computers were the only form of
computation \cite{1}. Starting from 1960s, digital computers drove them
out, mostly because analog computers are not general-purpose. Nowadays,
with the increased importance of computational speed and energy saving,
analog computers are actively studied again via quantum computation
\cite{2}, neuromorphic computation, and other approaches \cite{3}. None
of them will replace digital computers in the near future. However,
studying analog computers is important, since they can be combined with
digital ones, improving their speed and/or energy consumption; e.g., noisy, intermediate-scale quantum computers are considered from this
perspective \cite{4}. The major drawback of quantum computers is their
strong sensitivity to noise (decoherence) \cite{2,4}, which combines
with other drawbacks of quantum computers (e.g., non-scalability)
\cite{5}.

The unstructured database search is one of the main hallmarks of quantum
computation, because the unitary quantum computers can perform this task
much faster than classical digital computers
\cite{grover,farhi,review_grover}. In the simplest version, the
unstructured database consists of $M\gg 1$ elements, where $M-1$
elements are identical, and one is selected. Once the database is
unstructured, the classical digital computer should check one element
after another, which makes, on average, $(M+1)/2={\cal O}(M)$ checking
steps. A reformulation of this result highlights biophysical
applications of the unstructured database search, where a target
(selected) molecule should be found in a cell, or a target channel
should be found in a membrane \cite{bresloff,benichou_prl}. Assume that
the database consists of $M$ molecules (balls of size $a$) homogeneously
distributed in the cell; one of them is a selected target, which should
be reached by a random walker (e.g., a protein). The characteristic of
reaching the target is $\sim M/(nDa)={\cal O}(M)$, where $n$ is the
concentration of molecules, and $D$ is the diffusion constant of the
random walker in the cell \cite{bresloff}. The phenomenon of 
facilitated diffusion refers to introducing structural features into the
database \cite{bresloff}. 

A quantum, unitary computer completes searches the unstructured database
within $\sqrt{M}$ steps \cite{grover,farhi} (Grover's search), and this
is the best possible result for such computers
\cite{zalka,review_grover}; cf.~\cite{footnote1}. The analog
implementation of Grover's search looks for an $M$-level quantum system,
where one (selected) energy level has energy $\Delta E$, whereas all
other energy levels have energy zero \cite{farhi}. The quantum search
here amounts to running the Schroedinger dynamics for the $M$-level
system \cite{farhi}. Grover's search was generalized to classical waves
with the same conclusion on the search time $\sim\sqrt{M}$
\cite{grover_classical,patel,fourier}. 

We work out a classical, dissipative, long-range interacting many-body
model for unstructured database search, where the search time scales as
${\cal O}(M^{a})$ with $a<1/2$, i.e. it is better than Grover's search.
The advantage of this model is that it does not involve quantum
coherence, hence it is not susceptible to the main source of noise
(decoherence) that plagues standard quantum computers. Here, the
computation process amounts to spontaneous relaxation towards
equilibrium; i.e., there is no need to isolate the computer from its
environment and thereby maintain a unitary evolution, or any other
specific control operation. Thus, the search works via free energy
minimization. In that sense, it is similar to dissipative physical
devices performing various computational tasks; see \cite{yablo} for a
review.  

Our model for unstructured database search is based on the spherical
spin model, where the database is implemented via inter-spin
interactions with one selected (ferromagnetic) interaction; see below.
The final result is read off via magnetization measurements, which is
concentrated on the two spins that interact via the selected
interaction. These measurements take a shorter time than the relaxation, 
provided that the concentration is sufficiently strong. The drawback 
of this and all other {\it classical} search
models is that the database is implemented via ${\cal O}(\sqrt{M})$
classical spins. In the quantum situation, one needs ${\cal O}(\ln
M)$ quantum spins due to the tensor-product structure of composite quantum 
states \cite{grover,review_grover,farhi}. 

The spherical model relates to the Heisenberg
$O(n)$ spin model in the limit $n\gg 1$ \cite{thuemann}.  Due to its
flexibility and solvability, the spherical model was applied to a wide
range of problems, and led to a better understanding of new areas of
statistical physics: phase-transition \cite{review,baxter}; spin-glass
physics \cite{kosterlitz,barrat, cugliandolo_dean}; quantum critical
systems \cite{tm}; {\it etc}.  We show that it elucidates as well
capabilities of classical dissipative computers. 

{\it The database and its selected element.}
The Hamiltonian $H$ of the spherical model is defined over $N$ real variables (spherical spins) $S_\i$,
where $\i$ is the space-index:
\begin{align}
\label{1}
H=-\frac{1}{2}{\sum}_{\i\not=\j=1}^N\J_{\i\j}S_\i S_\j-\eps{\sum}_{\i=1}^Nh_{\i}S_\i,
\end{align}
where $\J_{\i\j}$ is the interaction matrix, and $\eps h_\i$ are external fields ($\eps>0$). 
Eq.~(\ref{1}) is supplemented with the mean spherical constraint ${\sum}_{\i=1}^N \overline{S_{\i}^2}= N$. It bounds 
energy $H$ and magnetization $\sum_{\i=1}^NS_{\i}$ and ensures well-defined equilibrium states.

For the interaction matrix $\J$ in (\ref{1}) we assume that one among its $M=N(N-1)/2$ elements is selected, while
all other elements are equivalent; e.g.,
\BEA
\label{z1}
\J_{12}=\J_{21}=R, ~~~
\J_{\i\not=\j, \i\geq 3}=\J_{\i\not=\j, \j\geq 3}=\frac{J}{N-3}, 
\EEA
where the spherical spins $S_1$ and $S_2$ interact via $R={\cal
O}(1)>0$, while all other interactions go via $\frac{J}{N-3}$ with
$J={\cal O}(1)$. This choice of $R$ and $J$ ensures that the energy $H$
is ${\cal O}(N)$ for $N\gg 1$.  A selected interaction means a selected
pair of spins; i.e., $S_1$ and $S_2$ for (\ref{z1}). The selected pair
in the Hamiltonian (\ref{1}) is unknown to us, though we know the values
of $R$ and $J$.  We want that in the equilibrium state generated by
(\ref{1}), the mean magnetization $\{\overline{S_{\i}}\}_{\i=1}^N$ concentrates on
$\i=1,2$. The maximal value allowed by the spherical constraint is
$\overline{S_1}=\overline{S_2}={\cal O}(\sqrt{N})$, and we want this
value to be much larger than the collective magnetization of all other
(nonselected) spins $\sum_{\i\geq 3}\overline{S_{\i}}$.  
The concentration is sufficiently large to ensure that the selected
spins can be found via ${\cal O}(\ln N)$ {\it finite-accuracy} magnetization measurements; see below and also section 1 of Supplementary Material (SM). 

Note that to get $\overline{S}_\i\not=0$, we need to have
external fields in (\ref{1}). These fields also should not know which
interaction is selected. The search time is the relaxation to the
equilibrium starting from some state that also does not know about the
selected interaction. We see below that there is a trade-off: a larger
$\overline{S_1}=\overline{S_2}$ requires longer relaxation time. But the
search time ${\cal O}(M^a)$ with $a<1/2$ is smaller than Grover's search
time. 

{\it Relaxation dynamics} of the model is described via over-damped
Langevin equations, where each spin $S_\i$ feels the conservative force
$-\partial_{S_{\i}}H$, friction force from the bath, and random noise \cite{landau,petr}.
The inertia $\propto \ddot{S}_{\i}$ is neglected assuming that the friction is strong. Recall that the Langevin equations are derived from 
system-bath Hamiltonian models, where the bath has an infinite number of degrees of freedom \cite{petr}. 
We use ket (bra) notations for column (row) vectors: $\langle
S|=(S_1,...,S_N)$, $\langle h|=(h_1,...,h_N)$, and $\langle
\i|=(0,..1..,0)$, where the single $1$ in $\langle \i|$ is located at
the $\i$'th place. Langevin's dynamics reads \cite{landau}:
\begin{align}
\label{d1}
&\frac{\d}{\d t}{|S(t)\rangle}=(\J-z(t){\cal I})|S(t)\rangle+\epsilon|h\rangle+|\eta(t)\rangle,\\
& \langle \eta|=(\eta_1,...,\eta_N),\quad \overline{\eta_\i(t)\eta_\j(t')}=2T\delta_{\i\j}\delta(t-t'),
\label{d2}
\end{align}
where $|\eta\rangle$ is the random white noise with zero mean, $\overline{|\eta\rangle}=0$, $T$ is the temperature,
$z(t)$ is the Lagrange multiplier ensuring the mean spherical constraint $\overline{\langle S|S\rangle}=N$,
and where the friction constant is 1. Eqs.~(\ref{d1}, \ref{d2}) is solved via the eigenresolution of the symmetrix matrix ${\cal J}$: 
\begin{align}
\label{dy0}
&{\cal J}|\lambda\rangle=\lambda|\lambda\rangle,\quad
\langle\lambda|\lambda'\rangle=\delta_{\lambda\lambda'}\quad \lambda, \lambda'=1,..,N,\\
\label{dy1}
&\langle \lambda |S(t)\rangle=e^{\lambda t-\hat z(t)}\langle \lambda |S(0)\rangle 
+\epsilon\langle\la |h\rangle e^{\lambda t-\hat z(t)}\times\nonumber\\
&\int_0^t\d s\, e^{-\lambda s+\hat z(s)}
+ e^{\lambda t-\hat z(t)}\int_0^t\d s\, e^{-\lambda s+\hat z(s)}\langle\la |\eta(s)\rangle,\\
\label{dy2}
&\hat z(t)=\int_0^t\d s\, z(s),\quad \overline{\langle \lambda |\eta(t)\rangle}=0,\\
&\overline{\langle \lambda |\eta(t)\rangle \langle \lambda' |\eta(s)\rangle}=2T\delta_{\lambda\lambda'}\delta(t-s).
\end{align}
$\langle \lambda |S(0)\rangle$ does not depend on $\langle\lambda |\eta(t)\rangle$.
The spherical constraint $\overline{\langle S|S\rangle}=\sum_{\lambda}\langle\lambda|S(t)\rangle^2=N$ produces from (\ref{dy1}) 
a non-linear integral equation for $e^{\hat z(t)}$, which is 
solved for a large $t$ via \cite{cugliandolo_dean}
\BEA
e^{\hat z(t)}\simeq e^{(\sigma+\lambda_1) t},\quad \sigma>0,
\label{dy9}
\EEA
where $\lambda_1$ is the largest eigenvalue of $\J$, and 
where the terms neglected in (\ref{dy9}) are exponentially small for long times. 
Eqs.~(\ref{dy1}, \ref{dy9}) show that each $\langle \lambda |S(t)\rangle$ relaxes with its own relaxation time, but
$\langle \lambda_1 |S(t)\rangle$ has the largest relaxation time $1/\sigma$.
Putting (\ref{dy9}) into (\ref{dy1}), using (\ref{dy2}), 
and taking the large $t$ limit, we get in the stationary state
\begin{align}
\label{7}
&1=\frac{T}{N}\sum_{\lambda} \frac{1}{\sigma+\lambda_1-\lambda }
+\frac{\eps^2}{N}\langle h|[\sigma+\lambda_1-\J]^{-2}|h\rangle,\\
&\overline{ S_{\i}}=\eps\langle\i|[\sigma+\lambda_1-\J]^{-1}|h\rangle,
\label{9}
\end{align}
where (\ref{7}) comes from $\overline{\langle S|S\rangle}=N$ and determines $\sigma>0$. 
Eq.~(\ref{9}) shows that $\overline{ S_{\i}}\to 0$ if $\epsilon\to 0$ for all other parameters being fixed.
This follows from $S_{\i}\to -S_{\i}$ symmetry of $H|_{\epsilon\to 0}$ in (\ref{1}).
Eqs.~(\ref{7}, \ref{9}) are found as well from the equilibrium Gibbs distribution based on (\ref{1}); see 
section 2 of SM. There $\sigma$ relates to the Lagrange multiplier
enforcing $\overline{\langle S|S\rangle}=N$.

\begin{figure}[ht]
\includegraphics[width=8cm]{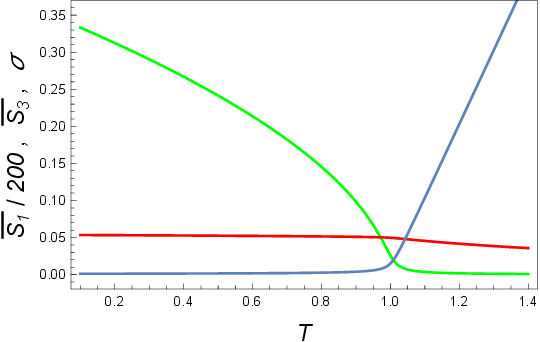}
\caption{For the spherical model (\ref{1}, \ref{7}, \ref{9}, \ref{gar1}--\ref{gar4}) 
we show the mean of the selected spin $\overline{S_1}/200$ (green), the mean of a nonselected
spin $\overline{S_3}$ (red) and the inverse relaxation time $\sigma$ as a function of temperature for $N=10^4$, $R=1$, $J=0.2$,
$\epsilon=0.05$ (weak fields). The external fields $\{h_\i\}_{\i=1}^N$ are random Gaussian variables with 
$w=1.54$, $h_1=0.33$, and $h_2=2.77$; see (\ref{ingva4}--\ref{ingva5}). 
$T=R$ is the phase-transition
point, where $\overline{S_1}$ starts to increase from a nearly zero value to $\overline{S_1}\propto \sqrt{N}$ for $T\to 0$. $\sigma$
monotonically decreases, assumes small values for $T<R$, but does not decrease much for $T<R$. $\overline{S_3}$ is nearly constant. 
For details of the plots see section 3 of SM. 
}
\label{fig1}
\end{figure}

{\it The stationary state.} Eqs.~(\ref{7}, \ref{9}) contain
the eigenresolution of $\J$, which reads from (\ref{1}, \ref{z1}):
\begin{align}
\label{ak1}
&\J={\sum}_{k=1}^3\lambda_k|\lambda_k\rangle\langle\lambda_k|-\frac{J}{N-3}\Pi_{N-3},\\
&\lambda_k=\frac{1}{2}\Big[
R+J\pm\sqrt{(R-J)^2 +\frac{8J^2(N-2)}{(N-3)^2}  }
\Big], \\
& \langle\lambda_k|=\frac{1}{\sqrt{2c^2(\lambda_k)+N-2}}[c(\lambda_k), c(\lambda_k), 1,1,...,1],\\
\label{ak5}
& c(\lambda)=\frac{J(N-2)}{(N-3)(\lambda-R)},\quad k=1,2,\\
& \lambda_3=-R,\qquad \langle \lambda_3|=\frac{1}{\sqrt{2}}[1,-1, 0,0,...,0],
\label{ak6}
\end{align}
and where $\Pi_{N-3}$ is a projector on the $N-3$ dimensional subspace: $[0,0,c_3,...,c_N]$ with $\sum_{k=3}^N c_k=0$: 
\BEA
\label{odzun}
&\Pi_{N-3}={\rm diag}[0,0,1,...,1]
-\frac{1}{N-2} |{\rm e}_3\rangle\langle{\rm e}_3|,~~\\
&\langle{\rm e}_3|=(0,0,1,...,1).
\EEA
The orthogonality and normalization of the eigenvectors are seen from (\ref{ak1}--\ref{ak5}).
Making natural assumptions
\BEA
N\gg 1,\qquad R>J,\quad 
J^2N^{-1}(R-J)^{-2}\ll 1, 
\label{asu}
\EEA
we get from (\ref{ak1}--\ref{ak5}):
\BEA
\label{gar1}
&&\lambda_1=R+\frac{2J^2}{N(R-J)}, \quad \lambda_2=J-\frac{2J^2}{N(R-J)},\\
&& \label{gar3}
\langle\lambda_1|= \frac{1}{\sqrt{2}}(1,1,\frac{2J}{N(R-J)},...,\frac{2J}{N(R-J)}),\\
&& \label{gar4}
\langle\lambda_2|= \frac{1}{\sqrt{N}}( \frac{J}{J-R},\frac{J}{J-R}, 1,...,1).
\EEA
Fig.~\ref{fig1} shows that for $N\gg 1$ and weak external fields
$\epsilon\ll 1$, the model described by (\ref{7}, \ref{9}, 
\ref{gar1}--\ref{gar4}) undergoes a phase-transition at $T=R$, where
the mean selected spins $\overline{S_1}$ and $\overline{S_2}$ start to
grow. For $T\leq R$, the inverse relaxation time $\sigma$ is small.
Both $\sigma$ and the mean non-selected spin $\overline{S_3}$ (and
$\overline{S_{\i>3}}$) are weakly-dependent on $T$ for $T\leq R$.  Hence,
we shall analytically study the regime of low temperatures $T\ll R$,
where $\overline{S_1}$ and $\overline{S_2}$ maximize. 

{\it Low temperatures.} Eq.~(\ref{7}) simplifies in the limit $T\to 0$
(i.e., $T\ll R$). Now the terms $\propto T$ in (\ref{7}) can be
neglected, and (\ref{7}) is solved by noting that the inverse relaxation
time $\sigma$ tends for $T\to 0$ to a constant, $\sigma={\cal
O}(\epsilon N^{-1/2})$, as confirmed in (\ref{khosrov}), and seen in
Fig.~\ref{fig1}. Hence we can neglect $\sigma$ everywhere besides the
term $\lambda=\lambda_1$ in (\ref{7}).  
The solution of (\ref{7}) reads
\BEA
\frac{\epsilon|\langle h|\lambda_1\rangle|}{\sigma\sqrt{N}}=\sqrt{
1-\frac{\epsilon^2 \frac{1}{N}\langle h|\lambda_2\rangle^2}{(R-J)^2}
-\frac{\epsilon^2 \frac{1}{N}\langle h|\Pi_{N-3}|h\rangle}{R^2} },
\label{khosrov}
\EEA
where the term $\propto \frac{1}{N}\langle h|\lambda_3\rangle^2={\cal
O}(\frac{1}{N})$ was neglected in (\ref{khosrov}); cf.~(\ref{ak6}). Note
that $\epsilon$ has to be sufficiently small for $\sigma>0$ in (\ref{khosrov}).  
Eq.~(\ref{9}) reads with the same approximation that led to
(\ref{khosrov}):
\begin{align}
\label{ingva1}
&\overline{ {S}_{\i}\,}=\sqrt{N}\langle\i|\lambda_1\rangle\,{\rm sign}[\langle h|\lambda_1\rangle]\times\nonumber\\
&\sqrt{
1-\frac{\epsilon^2 \frac{1}{N}\langle h|\lambda_2\rangle^2}{(R-J)^2}
-\frac{\epsilon^2 \frac{1}{N}\langle h|\Pi_{N-3}|h\rangle}{R^2}
}\nonumber\\
&+\frac{\epsilon\langle\i|\lambda_2\rangle\langle\lambda_2|h\rangle}{R-J }
+\frac{\epsilon\langle\i|\lambda_3\rangle\langle\lambda_3|h\rangle}{2R }
+\frac{\epsilon\langle \i|\Pi_{N-3}|h\rangle}{R}.
\end{align}
Note that $\overline{ {S}_{\i}\,}^2=\overline{ {S}^2_{\i}\,}$, as expected for $T=0$.

{\it Random external fields.} In (\ref{ingva1}) we can assume that the
external fields $\{h_\i\}_{\i=1}^N$ in (\ref{1}) are homogeneous:
$h_k=1$. This assumption leads to the same scaling $1/\sigma={\cal
O}(N)={\cal O}(\sqrt{M})$ of the relaxation time as in Grover's search;
see SM section 1.  Instead, we assume that each $h_k$ in $\langle
h|=(h_1,...,h_N)$ is Gaussian random variables with zero mean and
dispersion $1$. Recall that $\{h_\i\}_{\i=1}^N$ should not know about
selected spins 1 and 2. We show that random external fields do not
influence the magnetization of the selected spins, while the
magnetization of nonselected spins is suppressed. We consider only a
single realization of random variables $\{h_\i\}_{\i=1}^N$. 
Eqs.~(\ref{odzun}-\ref{gar4}) imply
\BEA
\label{ingva4}
&\langle {\rm e}_3|h\rangle=\sum_{k=3}^Nh_k=w\sqrt{N}, \\
&\frac{1}{N}\langle h|\Pi_{N-3}|h\rangle=1+{\cal O}(\frac{1}{\sqrt{N}}), \\
\label{ingva4.1}
&\langle h|\lambda_2\rangle=w+\frac{h_1+h_2}{\sqrt{N}}\frac{J}{J-R}=w+
{\cal O}(\frac{1}{\sqrt{N}}), \\ 
&\langle \i|\Pi_{N-3}|h\rangle=[h_{\i}-\frac{\langle {\rm e}_3|h\rangle}{N-2}]\delta_{\i\geq 3},
\label{ingva5}
\EEA
where $w$ is a Gaussian random variable with zero mean and dispersion $1$. 
Using (\ref{ingva4}--\ref{ingva5}) in (\ref{ingva1}) we get
\BEA
\label{ingva6}
&\overline{ {S}_{\i=1,2}\,}=\sqrt{\frac{N}{2}}{\rm sign}[h_1+h_2]
\sqrt{1-\frac{\epsilon^2}{R^2}}\\
&+\frac{\epsilon(h_1-h_2)(\delta_{\i 1}
-\delta_{\i 2})}{4R}-\frac{\epsilon}{\sqrt{N}}\,\frac{J(\delta_{\i 1}+ \delta_{\i 2})\langle\lambda_2|h\rangle}{(R-J)^2},\\
&\overline{ {S}_{\i\geq 3}\,}
=\sqrt{\frac{2}{N}}\,\frac{J}{R-J}{\rm sign}[h_1+h_2]
\sqrt{1-\frac{\epsilon^2}{R^2}}
\label{ingva07}\\
&+\frac{\epsilon\langle\lambda_2|h\rangle}{(R-J)\sqrt{N}}
+\frac{\epsilon[h_{\i}-\frac{\langle {\rm e}_3|h\rangle}{N-2}   ]}{R}.
\label{ingva7}
\EEA
The first term $\propto \sqrt{N}{\rm sign}[h_1+h_2]$ is the RHS of
(\ref{ingva6}) is the amplification of $\overline{ {S}_{\i=1,2}\,}$ due
to $R>J$. The sign (but not magnitude) of $\overline{ {S}_{\i=1,2}\,}$ is random and it
depends on the sign of the total field $h_1+h_2$ acting on the two
selected spins.  The term $\propto J$ in (\ref{ingva07})
is induced by the coupling to the selected spins. 

\comment{
The magnetization of
nonselected spins reads from (\ref{ingva4.1}, \ref{ingva5},
\ref{ingva7}):
\BEA
{\sum}_{\i\geq 3}\overline{ {S}_{\i}\,}=\frac{\sqrt{N}\epsilon\langle\lambda_2|h\rangle}{R-J}={\cal O}(\epsilon w\sqrt{N}).
\label{raor1}
\EEA
We get ${\sum}_{\i\geq 3}\overline{ {S}_{\i}\,}={\cal O}(\epsilon \sqrt{N})$, since $|w|\gg 1$ is highly improbable. 
}

We want to show that the selected spins $S_1$ and $S_2$ can be found in
${\cal O}(\ln [N])$ steps via measuring the mean magnetization of properly
defined subsets of $\{\i\geq 1\}$.  Eqs.~(\ref{ingva6}--\ref{ingva7})
are sufficient for this purpose, since they show that
$\overline{S_1}=\overline{S_2}={\cal O}(\sqrt{N})$ have the same order
of magnitude as $\sum_{\i\geq 3}\overline{S_{\i}}$.  But to make our
reasoning more transparent, we consider a sufficiently small $J$ so that
the term $\propto J$ in (\ref{ingva07}) can be neglected. Retaining this term
does not alter our main conclusions; see section 4 of SM.  We intend to
show that in any generic subset of spins containing spins $S_1$ and/or
$S_2$, the collective magnetization of the subset, is dominated by
$\overline{S_1}$ and/or $\overline{S_2}$, and is larger than in any
subset not containing them.  Here, generic means chosen in a way that 
is not correlated with the distribution of
$\{h_\i\}_{\i\geq 1}$. Let ${\cal K}$ be such a $K$-spin subset of
$\{\i\geq 3\}$.  We get from (\ref{ingva4.1}, \ref{ingva7}) 
(neglecting (\ref{ingva07})):
\BEA
\sum_{\i\in {\cal K}}\overline{ {S}_{\i}\,}=\frac{K\epsilon w}{(R-J)\sqrt{N}}+\frac{\epsilon}{R}
\frac{K}{\sqrt{N}}\Big(\sqrt{\frac{N}{K}}w_{\cal K}-w  \Big),
\label{raor2}
\EEA
where $w_{\cal K}=\frac{1}{\sqrt{K}}\sum_{\i\in {\cal K}} h_\i$ is a
Gaussian with mean zero and variance 1. $w$ and $w_{\cal K}$ are
correlated random variables. The maximal value of (\ref{raor2}) is
reached for $K\sim N$ and scales as ${\cal O}(\epsilon\sqrt{N})$. Comparing this with the dominant
term $\propto \sqrt{N}{\rm sign}[h_1+h_2]$ in (\ref{ingva6}), we see
that for $\epsilon={\cal O}(N^{-\zeta})$ (with $\zeta>0$),
the two selected spins $\overline{
{S}_{\i=1,2}\,}$ will dominate the behavior of any generic subset of $\{\i\geq
1\}$, which includes $\overline{ {S}_{\i=1}\,}$ or $\overline{
{S}_{\i=2}\,}$. If such a set does contain neither $S_1$, nor
$S_2$, then its mean magnetization will be much
smaller than for a set containing $S_1$ or $S_2$. 
Eq.~(\ref{raor2}) can be much larger than ${\cal O}({N}^{-\zeta+1/2})$ 
only due to e.g. $|w|\gg 1$. Such events are very rare and can be neglected.

Likewise, with $\epsilon={\cal O}(N^{-\zeta})$ we find from
(\ref{khosrov}, \ref{ingva4.1}, \ref{ingva5}) for the relaxation time: $1/\sigma={\cal
O}(N^{\zeta+0.5})$. For $0<\zeta<1/2$
this search time is shorter than the Grover search time ${\cal O}(N)$. 

The above concentration feature means that the selected spins $S_1$ and $S_2$ can be
found recursively: the total set of $\{\i\geq 1\}$ spins is divided into
two (not necessarily strictly equal) generic subsets and their (mean)
magnetizations re measured. If one of the subsets contains at least one
selected spin, its magnetization will be ${\cal O}(N^{0.5})$. Otherwise,
the magnetization will be at best ${\cal O}(N^{0.5-\zeta})$. The relative error of distinguishing such subsets from each other goes to zero for
$N\to\infty$. Then the target subset is divided into generic subsubsets, {\it
etc}. It should be clear that the number of steps for identifying both selected spins will then scale ${\cal O}(\ln[N])$, which is much
smaller than the relaxation time needed to amplify the selected spins. 

{\it Conclusion.} We studied the unstructured database search problem
via tools and models of classical statistical mechanics. This is a
relevant strategy in view of biophysical applications of the database
search: the protein folding problem, the performance optimization of
small heat-engines \cite{we_prl}, the target reaching problem in cell
biology \cite{bresloff} (c.f.~the introduction), {\it
etc}. We studied the classical two-body interacting spherical model
for illustrating the ideas of classical dissipative search, since this
model is solvable starting from the Langevin equations. Other models 
(e.g., multi-body interacting spherical model \cite{barrat,bolle})
might show a better performance as search engines.

The HESC of Armenia supported us under grants 24FP-1F030 and 21AG-1C038.
We thank Y. Mamasakhlisov and V. Stepanyan for discussions.

\clearpage

\onecolumngrid

\renewcommand{\theequation}{S\arabic{equation}}
\setcounter{equation}{0}
\renewcommand{\thefigure}{S\arabic{figure}}
\renewcommand{\thetable}{S\arabic{table}}

\begin{center}
{\bf\large Supplementary material for\\ ~ \\
{\sl Classical dissipative search of unstructured database} \\ ~ \\
by A.E. Allahverdyan and Y. Bisharyan
}
\end{center}

{\bf Description:} This Supplementary Material consists of the following
sections.  Section 1 discusses our basic assumptions: the accuracy of
magnetization measurements is finite in the thermodynamic limit. This is
exemplified for the spherical model under the homogeneous external
field. It shown in detail why the search time here has the same order of
magnitude as for Grover's search. Section 2 discusses the partition
function of the equilibrium spherical model. The presentation here is
standard and follows \cite{review,baxter}. The purpose of section 2 is
to confirm that Eqs.~(\ref{7}, \ref{9}) of the main text that were
deduced in the main text from the Langevin equations can also be recovered from the
equilibrium analysis.  Section 3 addresses the ferromagnetic
phase-transition and shows parameters and settings of Fig.~1 in the main
text. Section 4 provides the details of the derivation presented after
Eq.~(\ref{raor2}) of the main text. 
The formulas in this Supplementary Material are written as (\ref{muscat111}, \ref{muscat11}, ...).
Usual formulas refer to the main text and are indicated explicitly, e.g.: Eq.~(1) of the main text.

\section*{1. Finite-accuracy magnetization measurements and the spherical model under homogeneous external field}

We consider the homogeneous external field: $|h\rangle=(1,1,...,1)$. Using Eqs.~(\ref{ak1}--\ref{ak5}) of the main text,
$\langle R|h\rangle=0$ and $\Pi_{N-3}|h\rangle=0$, we get from Eq.~(\ref{ingva1}) of the main text:
\BEA
\label{muscat111}
&&\sigma=\frac{\epsilon}{\sqrt{N}}\,\frac{\sqrt{2}R}{R-J}\Big(1-\frac{\epsilon^2}{(R-J)^2}\Big)^{-1/2},\\
\label{muscat11}
&&\overline{ {S}_{\i}\,}=\sqrt{N}\Big[
\langle \i|\lambda_1\rangle
\Big( 1-\frac{\epsilon^2}{(R-J)^2} \Big)^{1/2}+\frac{\epsilon}{R-J}\langle \i|\lambda_2\rangle
\Big],\\
\label{ses0}
&&\overline{ {S}_{\i=1,2}\,}=\sqrt{\frac{N}{2}}\Big(1-\frac{\epsilon^2}{(R-J)^2}\Big)^{1/2}-\frac{\epsilon J}{(R-J)^2},\\
&& \overline{ {S}_{\i\geq 3}\,}= 
\frac{\epsilon}{R-J}+
\sqrt{\frac{2}{N}}\Big(1-\frac{\epsilon^2}{(R-J)^2}\Big)^{1/2}\frac{ J}{R-J},
\label{ses1}
\EEA
where $\langle \i|\lambda_1\rangle$ and $\langle \i|\lambda_2\rangle$ in (\ref{muscat111}) and (\ref{muscat11})
are deduced from Eq.~(\ref{gar3}) and Eq.~(\ref{gar4}) of the main text, respectively. 

Now (\ref{muscat111}) provides the inverse relaxation time, while
(\ref{muscat11}, \ref{ses0}, \ref{ses1}) describe magnetizations. 
The spatial distribution of $\overline{ {S}_{\i}\,}$
concentrates on those two spherical spins $\overline{ {S}_{1}\,}$ and
$\overline{ {S}_{2}\,}$ that interact via the biased bond with the
strength $R$. When this concentration is strong enough for our purposes?

We want to find the selected spins $S_1$ and $S_2$ by dividing the spins into two
equal-size groups. Assume for simplicity that the two selected spins are in the first
group. First, assume that $\epsilon={\cal O}(1)$ in (\ref{ses0}, \ref{ses1}).
Then the mean magnetization of the first and the second group are,
respectively [see (\ref{ses0}, \ref{ses1})], 
\BEA
\label{signals1}
\frac{\epsilon N}{R-J}+\sqrt{2N}\Big(1-\frac{\epsilon^2}{(R-J)^2}\Big)^{1/2}\quad {\rm and}\quad
\frac{\epsilon N}{R-J}.
\EEA
To distinguish these two signals from each other, the error (or the
inverse accuracy) of the measuring device should be smaller than the
relative difference between the two signals, i.e., smaller than ${\cal
O}(N^{-1/2})$, which is neglegible in the thermodynamic limit $N\gg 1$.
The same results is found when the selected spins are in different groups. 

Hence distinguishing between the signals (\ref{signals1}) demands excessively accurate
measurements. We exclude such measurements from our consideration by
demanding the measurement accuracy stays finite in the thermodynamic
limit $N\gg 1$.  This demand is completely standard and is present as
well in all unstructured database problems, which are supposed to
operate in the thermodynamic limit (i.e., for large databases). 

Instead of $\epsilon={\cal O}(1)$, we now assume in (\ref{ses0}, \ref{ses1})
that 
\BEA
\label{era}
\epsilon=\hat\epsilon N^{-1/2}, \quad \hat\epsilon={\cal O}(1).
\EEA
Instead of (\ref{signals1}), we now find from (\ref{ses0}, \ref{ses1})
for the mean magnetizations of the first and second groups:
\BEA
\label{signals2}
\sqrt{2N}+
\frac{\hat\epsilon \sqrt{N}+\sqrt{2N} J}{R-J} 
\quad {\rm and}\quad
\frac{\hat\epsilon \sqrt{N}+\sqrt{2N} J}{R-J}. 
\EEA
Now the relative difference between the signals is ${\cal O}(1)$, i.e.,
sufficiently accurate (but finite for $N\gg 1$) measurements can
distinguish between the two signals in (\ref{signals2}).  Note that this
conclusion is based on knowing sufficiently precisely the number of
spins in each group. The error in this knowledge should be much smaller
than $N/2$ (the number of elements in each group). 

Note from (\ref{muscat111}) that $\epsilon\propto N^{-1/2}$ means that
the relaxation time $1/\sigma$ is ${\cal O}(N)$, which is the Grover
regime.  We conclude that once finite-accuracy measurements are
employed, the homogeneous fields do not improve the quantum search time. 

Continuing along these lines, we shall reach groups whose size is smaller
than ${\cal O}(N)$. For such groups, we find instead of (\ref{signals1})
\BEA
\label{signals3}
\frac{\epsilon K}{R-J}+\sqrt{2N}\Big(1-\frac{\epsilon^2}{(R-J)^2}\Big)^{1/2}\quad {\rm and}\quad
\frac{\epsilon K}{R-J},
\EEA
where $K$ is the group size: $K\ll N$. Naturally, such two signals are much easier to distinguish than those in (\ref{signals1}).

To close this section, let us discuss why one cannot design an
analog search engine with the spherical model, where no inter-spin
couplings are involved. To this end, consider
\begin{align}
\label{ss1}
H=
-\eps{\sum}_{\i=1}^Nh_{\i}S_\i,
\end{align}
where $\eps h_\i$ are external fields ($\eps>0$), and where the spherical constraint $\sum_{\i=1}^N \overline{S^2_{\i}}=N$ is assumed. 
Note that the size of the database is now $N$, since the database is constructed via external fields.

Now Eqs.~(\ref{7}, \ref{9}) of the main text read:
\begin{align}
\label{ss7}
&1=\frac{T}{\sigma }
+\frac{\eps^2}{\sigma^2}\frac{1}{N}\langle h|h\rangle,\\
&\overline{ S_{\i}}=\frac{\eps h_{\i}}{\sigma}.
\label{ss9}
\end{align}
Eq.~(\ref{ss7}) is a quadratic equation for $1/\sigma$, which is solved for $T\to 0$, as 
\BEA
\label{ss9.1}
\frac{\eps}{\sigma}=\sqrt{\frac{N}{\langle h|h\rangle}}. 
\EEA
Hence,
we find from (\ref{ss9}) for $T\to 0$:
\BEA
\label{ss9.2}
\overline{ S_{\i}}=h_{\i}\sqrt{\frac{N}{\langle h|h\rangle}}, \qquad \frac{\overline{ S_{\i}}-\overline{ S_{\cal J}} }{\overline{ S_{\i}}}
=\frac{h_{\i}-h_{\cal J}}{h_{\i}}.
\EEA
It is seen from (\ref{ss9.2}) that the potential amplification factor $\sqrt{\frac{N}{\langle h|h\rangle}}$
is the same for all spins, and the relative spin differences are equal to those of the external fields. Since we
evaluate the mean magnetization of various subsets of $\{\i\geq 1\}$ based on the relative magnetization 
differences, the usage of the spherical model will not provide any advantage; see (\ref{ss9.2}).

\comment{
Now generically, the selected spin (for clarity, we assume that this is the first spin with $\i=1$) has 
$h_{1}\gg h_{\i\geq 2}$. To ensure the dominance of $\overline{ S_{1}}$ in the set of all spins, we should require at least
\BEA
\label{ss9.3}
\overline{ S_{1}}\sim \sum_{\i\geq 2}\overline{ S_{\i}}. 
\EEA
Comparing (\ref{ss9.2}) with (\ref{ss9.3}), we see that the difference $h_{1}\gg h_{\i\geq 2}$ should be pre-programmed in the choice 
of the external fields. Another way to state this point is to note that we should have 
$h_{1}={\cal O}(1)$ and $h_{\i\geq 2}={\cal O}(1)$, so that the difference 
$h_{1}-h_{\i\geq 2}>0$ is finite and does not depend on $N$. This precludes (\ref{ss9.3}) due to (\ref{ss9.2}).
}

\section*{2. Gibbsian equilibrium of the spherical model}

The statistical sum reads from Hamiltonian
\begin{align}
\label{s1}
H=-\frac{1}{2}{\sum}_{\i\not=\j=1}^N\J_{\i\j}S_\i S_\j-\eps{\sum}_{\i=1}^Nh_{\i}S_\i,
\end{align}
and the spherical constraint $\sum_{\i=1}^NS_{\i}^2=N$:
\BEA
\label{3}
&&Z=\int\prod_{\i=1}^N\d S_\i\,  e^{-\beta H[S]}\,\delta({\sum}_{\i=1}^N S_{\i}^2- N)=\int\prod_{\i=1}^N\d S_\i
\int_{\Re u-i\infty}^{\Re u+i\infty}\frac{\d u}{2\pi i}
e^{Nu-\sum_{\i\j}(u\delta_{\i\j}-\frac{\beta \J_{\i\j}}{2})S_\i S_\j +\beta\epsilon \sum_{\i}h_\i S_\i
}\\
&&= \pi^{N/2}\int_{\Re u-i\infty}^{\Re u+i\infty}\frac{\d u}{2\pi i}\, e^{Ng(u)},
\qquad \Re u> \frac{\beta\lambda_{\rm max}}{2},
\label{4} \\
\label{5}
&& g(u)\equiv u-\frac{1}{2N}\sum_{\lambda} \ln (u-\frac{\beta\lambda}{2})+\frac{\eps^2\beta^2}{4N}
\langle h|[u\uno-\frac{\beta}{2}\J]^{-1}|h\rangle,
\EEA
where $\uno=\{\delta_{\i\j}\}$ is the unity matrix, and 
where in the integration over $u$ we demand $\Re u> \frac{\beta\lambda_{\rm max}}{2}$ 
with $\lambda_{\rm max}$ being the maximal eigenvalue of $\J_{\i\j}$. $\{\lambda\}$ are eigenvalues of matrix $\J$. When going from
(\ref{3}) to (\ref{4}) we changed variables as 
\BEA
|S'\rangle=|S\rangle-\frac{\beta\eps}{2} [u-\frac{\beta}{2}\J]^{-1}|h\rangle,
\label{6}
\EEA
and took the Gaussian integrals over $S'$. Condition $\Re u> \frac{\beta\lambda_{\rm max}}{2}$ is needed for ensuring the
convergence of these integrals.

The integral in (\ref{4}) is taken via the saddle-point method for $N\gg 1$. The saddle-point value is determined from
\BEA
\label{7s}
&&\frac{\partial g}{\partial u}=1-\frac{1}{2N}\sum_{\lambda} \frac{1}{u-\frac{\beta\lambda}{2}}
-\frac{\eps^2\beta^2}{4N}\langle h|[u-\frac{\beta}{2}\J]^{-2}|h\rangle=0,\\
&&\frac{\partial^2 g}{\partial u^2}>0,
\label{7.1s}
\EEA
where the last relation means that for real value of $u$, $g(u)$ has to have a minimum, and then for imaginary $u$ it will have a maximum, and
lead to the saddle-point method for $N\gg 1$. Note that (\ref{5}) implies 
\BEA
\label{7.2s}
\frac{1}{\beta}\,\frac{\partial g}{\partial  \eps}=\frac{1}{N}\sum_{\i=1}^N \overline{ S_\i} h_\i,
\EEA
where $\overline{ S_\i}$ means averaging over the Gibbs distribution; see (\ref{3}). Now (\ref{7s}) shows that 
$\frac{\partial g}{\partial  \eps}=\frac{\partial g}{\partial  \eps}\Big|_{u}$, i.e. from (\ref{7.2s}) we find
\BEA
\frac{1}{N}\sum_{\i=1}^N \overline{ S_\i} h_\i=
\frac{\eps\beta}{2N}\langle h|[u\uno-\frac{\beta}{2}\J]^{-1}|h\rangle.
\label{8}
\EEA

Eq.~(\ref{8}) can be also obtained from (\ref{6}). After applying to this equation $\overline{ |S'\rangle}=0$ 
(only Gaussian integrals are involved that symmetric with respect to $S'_\i\to -S'_\i$), we get  
\BEA
\overline{| S\rangle}=\frac{\beta\eps}{2} [u\uno-\frac{\beta}{2}\J]^{-1}|h\rangle=
\frac{\beta\eps}{2}\sum_{\lambda}\frac{\langle \lambda|h\rangle}{u-\frac{\beta\lambda}{2}}\, |\lambda\rangle,
\label{9.a}
\EEA
where $u$ is determined from (\ref{7s}).

Let $\hat S_\lambda$ be the normal modes obtained after the diagonalizing orthogonal transformation
in (\ref{3}, \ref{6}): 
\BEA
S'_\i=\sum_\lambda\langle\i|\lambda\rangle\hat{S}_\lambda,\qquad
\hat S_\lambda=\sum_{\i}\langle\lambda|\i\rangle S'_\i
\EEA
Using (\ref{3}) we get
\BEA
\label{de}
\overline{\hat S_\lambda^2\,}=\frac{1}{2(u-\frac{\beta\lambda}{2})},\qquad 
\overline{ {S'}_\i^2\,}=\frac{1}{2}\sum_\lambda \frac{\langle\lambda|\i\rangle^2 }{u-\frac{\beta\lambda}{2}},\qquad
\overline{ {S'}_\i\,}=0.
\EEA 

Eqs.~(\ref{de}, \ref{9}, \ref{6}) imply
\BEA
\label{der}
\overline{ {S}_\i^2\,}=\frac{1}{2}\sum_\lambda \frac{\langle\lambda|\i\rangle^2 }{u-\frac{\beta\lambda}{2}} 
+\frac{\beta^2\eps^2}{4} \langle \i|[u-\frac{\beta}{2}\J]^{-1}|h\rangle^2.
\EEA 
Now summing (\ref{der}) over $\i$ we revert to the spherical constraint (\ref{7s}), which clarifies its physical meaning.

In the same way, we calculate the mean energy from (\ref{1}, \ref{5}, \ref{7s}):
\BEA
\frac{1}{N}\overline{H}=\frac{T(1-2u)}{2}-\frac{\eps^2\beta}{4N}\langle h|[u-\frac{\beta}{2}\J]^{-1}|h\rangle.
\EEA
The general concept of phase-transitions in the spherical model is that
for sufficiently low temperatures, the saddle-point solution $u$ in
(\ref{7s}) approaches for $\eps\to 0$ to its limiting value
$\frac{\beta\lambda_{\rm max}}{2}$. Further clarifications of the
phase-transition scenario demand understanding the eigen-structure of
$\J$. On regular lattices, the eigenvalues of $\J$ converge (in the
thermodynamic limit $N\gg 1$) to a well-defined distribution
\cite{review,baxter}. Hence, (\ref{7s}, \ref{8}) do not depend on $N$ and
can be studied directly in the thermodynamic limit \cite{review,baxter}. 

The parameter (Lagrange multiplier) $u$ obeying (\ref{7s}) relates to the parameter $\sigma$ used in the main text:
\BEA
\sigma=2uT-\lambda_1,
\EEA
where $\lambda_1$ is the maximal eigenvalue of the matrix $\J$ in (\ref{3}).

\section*{3. Finite temperatures and phase-transition in the spherical model}

Here we explain details of Fig.~\ref{fig1}.
Let us first note more detailed versions of Eqs.~(\ref{ingva4}--\ref{ingva5}) of the main text:
\BEA
\label{ingvas4}
&\langle {\rm e}_3|h\rangle=\sum_{k=3}^Nh_k=w\sqrt{N}, \\
&\frac{1}{N}\langle h|\Pi_{N-3}|h\rangle=\frac{1}{N}\sum_{k=3}^{N}h^2_k-\frac{\langle {\rm e}_3|h\rangle^2}{N(N-2)}
=1+\frac{1}{N}\sum_{k=3}^N (h_k^2-1) +{\cal O}(\frac{1}{N})=1+\frac{w_1}{\sqrt{N}}+{\cal O}(\frac{1}{N}),\\
\label{ingvas4.0}
&\langle h|\lambda_1\rangle=\frac{h_1+h_2}{\sqrt{2}} +\frac{2Jw}{\sqrt{N}(R-J)}=
\frac{h_1+h_2}{\sqrt{2}} + {\cal O}(\frac{1}{\sqrt{N}}), \\ 
\label{ingvas4.1}
&\langle h|\lambda_2\rangle=w+\frac{h_1+h_2}{\sqrt{N}}\frac{J}{J-R}=w+
{\cal O}(\frac{1}{\sqrt{N}}), \\ 
\label{ingvas4.2}
&\langle h|\lambda_3\rangle=\frac{h_1-h_2}{\sqrt{2}},\\
&\langle \i|\Pi_{N-3}|h\rangle=[h_{\i}-\frac{\langle {\rm e}_3|h\rangle}{N-2}]\delta_{\i\geq 3},
\label{ingvas5}
\EEA
where $w$ is a Gaussian random variable with zero mean and dispersion $1$; for $N\gg 1$, $w_1$ also converges to 
a Gaussian random variable with zero mean and dispersion $1$. 

We obtain from (\ref{ingvas4}--\ref{ingvas5}) and Eqs.~(\ref{7}, \ref{9}) of the main text:
\BEA
\label{ra1}
&
0=-1+\frac{T}{N}\Big[
\frac{1}{\sigma}+\frac{1}{\sigma+R-J}+\frac{1}{\sigma+2R}+\frac{N}{\sigma+R}
\Big]
+\frac{\epsilon^2}{N}\Big[
\frac{\langle h|\lambda_1\rangle^2}{\sigma^2}+\frac{\langle h|\lambda_2\rangle^2}{(\sigma+R-J)^2}
+\frac{\langle h|\lambda_3\rangle^2}{(\sigma+2R)^2}+\frac{N+w_1\sqrt{N}}{(\sigma+R)^2}
\Big],\\
\label{ra2}
&
\overline{S_1}=\frac{\epsilon\langle h|\lambda_1\rangle}{\sqrt{2}\sigma}
+ \frac{\epsilon J\langle h|\lambda_2\rangle}{\sqrt{N}(J-R)(\sigma+R-J)}
+\frac{\epsilon \langle h|\lambda_3\rangle}{\sqrt{2}(\sigma+2R)},\\
\label{ra3}
&
\overline{S_3}=\frac{\sqrt{2}\epsilon J\langle h|\lambda_1\rangle}{N(R-J)\sigma}
+ \frac{\epsilon \langle h|\lambda_2\rangle}{\sqrt{N}(\sigma+R-J)}+\frac{\epsilon(h_3-\frac{w}{\sqrt{N}})}{\sigma+R}
\EEA
Fig.~\ref{fig1} is contructed via (\ref{ra1}--\ref{ra3}), where we have put $w_1\to 0$ for simplicity.

\section*{4. Further details on the derivation after Eq.~(\ref{raor2}) of the main text}

We noted after Eqs.~(\ref{ingva6}--\ref{ingva7}) of the main text that
$\overline{S_1}={\cal O}(\sqrt{N})$ and $\overline{S_2}={\cal
O}(\sqrt{N})$.  After Eq.~(\ref{raor2}) of the main text we argued that
provided that $\epsilon={\cal O}(N^{-\zeta})$ with $0<\zeta<1/2$, we
ensure the relaxation time $1/\sigma={\cal O}(N^{\zeta+0.5})$, and we
achieve the mean magnetization ${\cal O}(N^{0.5-\zeta})$ for subsets of
$\{\i\geq 1\}$ that contain ${\cal O}(N)$ spins and do not contain the
two selected spins $S_1$ and $S_2$. However, these conclusions was
obtained after neglecting the term given by Eq.~(\ref{ingva07}) of the
main text. Now we retain this term.  The reasoning here is similar to
that in section 1. We repeat it for clarity and completeness. 

Let ${\cal K}$ be a generic $K$-spin subset of $\{\i\geq 3\}$, i.e.,
${\cal K}$ does not contain the selected spins $S_1$ and $S_2$. Here
generic means chosen in a way that is not correlated with the
distribution of $\{h_\i\}_{\i\geq 1}$. We get from Eqs.~(\ref{ingva4.1},
\ref{ingva7}) of the main text:
\BEA
\label{sraor1}
\sum_{\i\in {\cal K}}\overline{ {S}_{\i}\,}= f({\cal K})
+\sqrt{\frac{2}{N}}\,\frac{JK}{R-J}{\rm sign}[h_1+h_2]
\sqrt{1-\frac{\epsilon^2}{R^2}}, \\
f({\cal K})\equiv\frac{K\epsilon w}{(R-J)\sqrt{N}}+\frac{\epsilon}{R}
\frac{K}{\sqrt{N}}\Big(\sqrt{\frac{N}{K}}w_{\cal K}-w  \Big), \quad w_{\cal K}=\frac{1}{\sqrt{K}}\sum_{\i\in {\cal K}} h_\i.
\label{sraor2}
\EEA
Likewise, let ${\cal L}$ be a generic $L$-spin subset of $\{\i\geq 1\}$ that contains 
the selected spin $S_1$ and does not contain $S_2$. (Subsets that contain both 
selected spins are treated in a similar way.) We have from Eq.~(\ref{ingva6}) of the main text:
\BEA
\sum_{\i\in {\cal L}}\overline{ {S}_{\i}\,}
= f({\cal L})
+ \sqrt{\frac{N}{2}}{\rm sign}[h_1+h_2]
\sqrt{1-\frac{\epsilon^2}{R^2}}
+\frac{\epsilon(h_1-h_2)}{4R}+{\cal O}\Big(\frac{\epsilon}{\sqrt{N}}\Big)\nonumber\\
+\sqrt{\frac{2}{N}}\,\frac{J(L-1)}{R-J}{\rm sign}[h_1+h_2] \sqrt{1-\frac{\epsilon^2}{R^2}}.
\label{sraor3}
\EEA
For $\epsilon={\cal O}(N^{-\zeta})$ with $0<\zeta<1/2$ we have from (\ref{sraor2})
\BEA
\label{sraor4}
f({\cal K})= {\cal O}(KN^{-0.5-\zeta})+{\cal O}(N^{-\zeta}).
\EEA
It is seen from (\ref{sraor1}, \ref{sraor2}, \ref{sraor3})
that $\sum_{\i\in {\cal L}}\overline{ {S}_{\i}\,}\gg \sum_{\i\in {\cal K}}\overline{ {S}_{\i}\,}$ unless $K={\cal O}(N)$.
If this condition holds, the relative difference 
between two experimental signals $\sum_{\i\in {\cal L}}\overline{ {S}_{\i}\,}$ and $\sum_{\i\in {\cal K}}\overline{ {S}_{\i}\,}$ reads
\BEA
\label{sraor5}
\frac{\sum_{\i\in {\cal L}}\overline{ {S}_{\i}\,}-\sum_{\i\in {\cal K}}\overline{ {S}_{\i}\,}}{\sum_{\i\in {\cal L}}\overline{ {S}_{\i}\,}}
=1-\frac{2J(K-L+1)}{(R-J)N}.
\EEA
Eq.~(\ref{sraor5}) is of order of 1 for $K={\cal O}(N)$. It depends only on known quantities, i.e. sufficiently accurate (but with a finite
accuracy for $N\gg 1$) measurements can distinguish between 
$\sum_{\i\in {\cal L}}\overline{ {S}_{\i}\,}$ and $\sum_{\i\in {\cal K}}\overline{ {S}_{\i}\,}$.


\begin{thebibliography}{99}
\bibitem{footnote1} Ref.~\cite{6} proposed a framework for quantum
dissipative unstructured database search, where even smaller search times are
possible. However, the framework of \cite{6} is phenomenological and
does not involve explicit microscopic mechanisms. This major drawback is corrected here on the classical level.
\end{thebibliography}
\end{document}